\documentstyle[11pt]{article}
\author
{W.~X.~Ma
\thanks {On leave of absence from Institute of
Mathematics, Fudan University, Shanghai 200433, China}$\ $
  and B.~Fuchssteiner\\
FB17, Mathematik-Informatik, Universitaet Paderborn,
\\ 33098 Paderborn, Germany }
\title
{Explicit and Exact Solutions to a Kolmogorov-Petrovskii-Piskunov Equation}
\date{\nonumber}
\begin{document}
\maketitle
\begin{abstract}
Some explicit traveling wave
solutions to a Kolmogorov-Petrovskii-Piskunov equation
 are presented through two ans\"atze. By a Cole-Hopf
transformation, this
 Kolmogorov-Petrovskii-Piskunov equation is also  written
as a bilinear equation
 and further two solutions to describe nonlinear interaction
of traveling waves are generated. B\"acklund transformations of the linear
form and some special cases are considered.
\end{abstract}



\section{Introduction}

\def \la {\lambda}
\def \be {\beta}
\def \al{\alpha}
\def \del{\delta}
\def \Del{\Delta}
\def \al{\alpha}
\def \vare{\varepsilon}

\def \part {\partial}

The Cauchy problem of the Kolmogorov-Petrovskii-Piskunov equation
 \cite{KPP}
\[\left \{\begin{array}{l} u_t-u_{xx}=f(u),\ f \ \textrm {nonlinear},\
f(0)=0,
\\u(x,0)=\phi (x),\ x\in R^1\end{array}
\right.\]
has been extensively investigated both by analytic
techniques \cite{Aronson}, \cite{Kametaka}, and by probabilistic
methods \cite{Mckean}, \cite{Bramson}, and the existence of traveling
wave solutions with various velocities has been also proved.
A special case is the Fisher equation
\[ u_t-u_{xx}=u(1-u),\]
which was originally proposed \cite{Fisher} as a model for the propagation
 of a favored gene.
An explicit and exact solitary  solution of the Fisher equation
 may be presented by
Painlev\'e analysis \cite{Ablowitz}.
As an example of the above Kolmogorov-Petrovskii-Piskunov equation,
Kametaka \cite{Kametaka} considered the Cauchy problem of a generalized
Fisher equation
\[ \left\{\begin{array}{l} u_t-u_{xx}=\la _0^2u(1-u^n),
\\ u(x,0)=\{1+(2^{n/2}-1)\textrm{e}^{-(n/2)\sigma x}\}^{-2/n},\ x\in R^1,
\end{array}\right.\]
where $\la _0>0, \,n\in N,\, \sigma >0,$ and gave an explicit solution
\[
 u(x,t)=\left[1+(2^{n/2}-1)\textrm{exp}\{-\frac n 2 \sigma _1(x+2\la _1t)\}
\right]^{-2/n}
\]
when
\[ \sigma =\sigma _1=\la _1 -\sqrt{\la _1 ^2-\la _0^2},\ \la _1=\frac12
\left\{(\frac n2
+1)^{1/2}+(\frac n2 +1)^{-1/2}\right\} \la _0.\]
Abdelkader \cite{Abdelkader} and Wang \cite{Wang}
extended the integer $n$ to a real number
$\al $ satisfying $\al \ge 1$
and Wang obtained a class of explicit traveling wave solutions
successfully by introducing a special
nonlinear transformation.

In this paper, we consider the following Kolmogorov-Petrovskii-Piskunov
equation
\begin{equation}u_t-u_{xx}+\mu u+\nu u^2+\del u^3=0\label{KPP}
\end{equation}
where $\mu ,\nu, \del $ are three real constants.
Some special cases with $\mu +\nu +\del =0$
of this equation have been studied (see for instance
\cite{Cohen},
\cite{Kawahara}, \cite{Hereman} and \cite{Krishnan}).
Our purpose is to look for explicit and exact solutions for the general
case of (\ref{KPP}). Note that a more general equation
\[ u_{t'}-\al u_{x'x'}+\mu u+\nu u^2+\del u^3=0,\ \al \ne0\]
may be mapped into the Kolmogorov-Petrovskii-Piskunov
equation (\ref{KPP}) by a time scaling $t'\to \al t$ and therefore the
 Kolmogorov-Petrovskii-Piskunov
equation (\ref{KPP}) is no loss of generality  .
In Section 2, we analyze the possibilities of solutions which correspond
to two Riccati equations and further give explicitly a number of exact
solutions
to (\ref{KPP}). The essential point is to break the nonlinear
equation (\ref{KPP}) into two smaller problems and
 then to solve these two smaller
problems.  In Section 3, we change the
 Kolmogorov-Petrovskii-Piskunov (\ref {KPP}) into a bilinear equation,
 like the
Hirota bilinear one, by making
 the well known Cole-Hopf transformation,
and present two explicit solutions to describe
nonlinear coalescence of traveling wave solutions. In Section 4, B\"accklund
transformations of the linear form are discussed  along
with some  explicit relations of B\"acklund transformations for
the obtained solutions. In Section 5, more
discussion is given.

\section{Traveling wave solutions}

We consider traveling wave solutions to
the Kolmogorov-Petrovskii-Piskunov (\ref {KPP})
\[ u(x,t)=u(\xi )=u(kx-\omega t),\]
where the wavenumber $k$ and the frequency $\omega$ are required to be nonzero
for generating nontrivial solutions.
The resulting ordinary differential equation from (\ref{KPP}) reads as
\begin {equation}
-\omega u_\xi -k^2 u_{\xi\xi}+\mu u+\nu u^2+\del u^3=0.
\label{basic}\end {equation}
In the following we generate
traveling wave solutions to (\ref{KPP}) starting with
two ans\"atze.

First we make the ansatz for (\ref {basic})
\begin {equation}
u(\xi )=\sum_{i= 0}^Ma_iv^i=\sum_{i= 0}^Ma_i\left(v(\xi )\right)^i,
\ v_\xi =\vare (1-v^2),\ \vare =\pm 1.
\label{Ansatz1}
\end {equation}
It is easy to show that $M$ must be 1 if the functions
$1, v, v^2,\cdots, v^m\, (m\in N)$
are linear independent and $\del \ne 0$. So without loss of
generality  , we may choose
\[ u=a_0+a_1v,\]
and thus
\[ \begin {array} {l} u_\xi=\vare a_1-\vare a_1 v^2,\\
u_{\xi\xi}=-2a_1v+2a_1v^3.\end{array}\]
The substitution into (\ref{basic})
yields the following conditions for determining $a_0,a_1,k,\omega $:
\begin {equation}
\left\{\begin{array}{l}
-\vare \omega a_1+\mu a_0+\nu a_0^2 +\del a_0^3=0,\\
2k^2a_1+\mu a_1+2\nu a_0a_1 +3\del a_0^2 a_1=0,\\
\vare \omega a_1 +\nu a_1^2 +3 \del a_0 a_1^2=0,\\
-2k^2a_1+\del a_1^3=0.
\end{array}\right.\label {condition1}
\end {equation}
We need to assume $a_1\ne
0$ for nontrivial solutions, and thereby we get
\begin {eqnarray}& &
k^2=\frac12 \del a_1^2,\ \omega =-\vare (\nu a_1+3\del a_0a_1),\label{1}
\\ & &a_1^2=-\frac  {\mu}{\del } -\frac {2\nu}{\del }a_0-3a_0^2,\label{2}
\\ & &f(a_0):=\frac {\mu \nu }{\del }+2(\frac {\nu ^2}\del +\mu )a_0+8\nu
 a_0^2+8\del a_0^3=0,\label{cubic}
\end {eqnarray}
in which (\ref{1}) shows that $\del >0$. The equation (\ref{cubic}) is a cubic
equation. Krishnan \cite{Krishnan} analyzed the case of (\ref {cubic})
with $\mu =a,\, \nu =-(a+1),\, \del =1$,
but he failed to give any solution. Actually, this equation has a solution $
a_0=-\frac \nu {2\del }$ by inspection.
We can accordingly decompose
\[ f(a_0)=(a_0+\frac \nu {2\del })(8\del a_0^2+4\nu a_0  +2\mu),\]
from which we acquire
three roots
\begin {equation}
a_{01}=-\frac \nu {2\del },\ a_{02,3}=\frac {-\nu \pm \sqrt{\Del }
}{4\del },\ \Del =\nu ^2-4\mu \del .
\end {equation}
Of course, we can also solve the cubic equation (\ref{cubic}) by
computer algebra tools, for example, Mathematica, Maple and MuPAD.
Further we can obtain by (\ref {2}) and (\ref {1})
\begin {eqnarray}
&& a_{11}=\vare _1\frac {\sqrt{\Del }}{2\del },\
 a_{12,3}=\vare _{2,3}a_{02,3} ;
\\ & &k_1=\vare _4 \frac {\sqrt{\Del }}{2\sqrt {2\del }},\
k_2=\vare _5 \frac {\nu -\sqrt{\Del }}{4\sqrt {2\del }},\
k_3=\vare _6 \frac {\nu +\sqrt{\Del }}{4\sqrt {2\del }};
\\ & &\omega=\vare \vare _1\frac {\nu \sqrt{\Del }}{4\del },\
 \omega  _{2,3}=-\vare \vare _{2,3}\left[\frac{(\nu \mp \sqrt
 {\Del })^2}{16\del
 }-\frac \nu 2\right].
\end {eqnarray}
Now we may conclude that only when $\del >0,\, \Del \ge 0$, there are
real solutions $a_0,a_1,k,\omega$
and
 we acquire
the following three  exact solutions for the
Kolmogorov-Petrovskii-Piskunov equation (\ref {KPP}) with
 $\del >0,\, \Del \ge 0$:
\begin {eqnarray}
& & u_1(x,t)=u_1(x,t;\vare _1,\vare _4)=a_{01}+a_{11}v(k_1x-\omega _1t)
\nonumber
\\
& & =-\frac \nu {2\del }+\vare _1\frac {\sqrt{\Del }}{2\del }
v(\vare _4\frac {\sqrt {\Del }}{2\sqrt {2\del }}x-\vare _1 \frac
{\nu \sqrt {\Del }}{ 4\del }t), \label{solu1}
\\& &u_2(x,t)=u_2(x,t;\vare _2,\vare _5)=a_{02}+a_{12}v(k_2x-\omega _2t)
\nonumber\\
  & & =\frac {-\nu +\sqrt {\Del }} {4\del }+\vare _2\frac {-\nu +
\sqrt{\Del }}{4\del }
v(\vare _5\frac {\nu -\sqrt {\Del }}{4\sqrt {2\del }}x+\vare _2 [\frac
{(\nu -\sqrt {\Del })^2}{ 16\del }-\frac \mu 2]t), \label{solu2}
\\& &u_3(x,t)=u_3(x,t;\vare _3,\vare _6)=a_{03}+a_{13}v(k_3x-\omega _3t)
\nonumber\\
 & & =\frac {-\nu -\sqrt {\Del }} {4\del }+\vare _3\frac {-\nu -
\sqrt{\Del }}{4\del }
v(\vare _6\frac {\nu +\sqrt {\Del }}{4\sqrt {2\del }}x+\vare _3 [\frac
{(\nu +\sqrt {\Del })^2}{ 16\del }-\frac \mu 2]t),\qquad\quad
\label{solu3}
\end {eqnarray}
where $v_\xi=1-v^2,\ \vare _i=\pm 1,\, 1\le i\le 6.$ In the above solutions,
we cancel the case of $\vare =-1$ due to the same solutions.

Notice that the Riccati equation
 $v_\xi =a (1-v^2)\ (a\in R^1)$ has a general solution
\begin{equation}
v=v(\xi )=\frac{A-B\textrm{e}^{-2a \xi }}{A+B\textrm{e}^{-2a \xi }}
=\left\{\begin{array}{cl}
1& \textrm{for}\ B=0,\\-1& \textrm{for}\ A=0,\\ \textrm{tanh}\left(
a \xi -\frac12 \textrm{ln}(\frac BA )\right)&\textrm{for}\ AB>0,\\
\textrm{coth}\left(
a \xi -\frac12 \textrm{ln}(-\frac B A )\right)&\textrm{for}\ AB<0,
\end{array}\right.
\end{equation}
where $A,B$ are arbitrary constants satisfying $A^2+B^2\ne 0$.
This solution may be obtained by three tricks: a M\"obius transformation,
a Cole-Hopf transformation or a relation \cite {Kamke}
\[\frac{ (v_1-v_2)(v_3-v_4)}{(v_1-v_3)(v_2-v_4)}=C,\
C=\textrm{const.}\]
of  the solutions $v_i,\,1\le i\le 4$, beginning with three known  solutions
 $ 1, -1, \textrm{tanh}(\xi )$.
{}From $v(\xi )=\pm 1 , $ (\ref{solu1}), (\ref{solu2}) and (\ref{solu3})
result in three constant solutions of (\ref{KPP}), but
after choosing
\[v(\xi )=\textrm{tanh}(\xi +\xi _0),\ v(\xi )=\textrm{coth}(\xi +\xi _0)\
(\xi_0\ \textrm{ arbitrary}),\]
 (\ref{solu1}), (\ref{solu2}) and (\ref{solu3})
yield non trivial solutions:
three explicit traveling front solutions and three explicit singular
traveling solutions, respectively.

Secondly we make another ansatz for (\ref {basic})
\begin {equation}
u(\xi )=\sum_{i= 0}^Mb_iv^i=\sum_{i= 0}^Mb_i\left(v(\xi )\right)^i,
\ v_\xi =\vare (1+v^2),\ \vare =\pm 1.
\label{Ansatz2}
\end {equation}
Similarly $M$ must equal 1 if the functions
$1,v,v^2,\cdots, v^m\, (m\in N)$ and $\del \ne 0$. So without
 loss of generality  ,
we can choose
\[ u=b_0+b_1v,\]
and further we find
\[ \begin {array} {l} u_\xi=\vare b_1+\vare b_1 v^2,\\
u_{\xi\xi}=2b_1v+2b_1v^3.\end{array}\]
The substitution into (\ref{basic})
engenders the following conditions on $b_0,b_1,k,\omega $:
\begin {equation}
\left\{\begin{array}{l}
-\vare \omega b_1+\mu b_0+\nu b_0^2 +\del b_0^3=0,\\
-2k^2b_1+\mu b_1+2\nu b_0b_1 +3\del b_0^2 b_1=0,\\
-\vare \omega b_1 +\nu b_1^2 +3 \del b_0 b_1^2=0,\\
-2k^2b_1+\del b_1^3=0.
\end{array}\right.\label {condition2}
\end {equation}
Note that there are only
two terms in (\ref {condition1}) and (\ref {condition2})
have opposite signs. In an analogous way,
we can prove that only when $\del >0,\,\Del \le 0$, there exists a set of
real nonzero solutions $
b_0,b_1, k, \omega$, which may be worked out
\begin {eqnarray}b_0=-\frac \nu {2\del },\ b_1=\vare _1\frac
{\sqrt {-\Del }}{2\del },\ k=\vare _2
\frac {\sqrt {-\Del}}{2\sqrt{2\del} },\
\omega =-\vare \vare _1 \frac {\nu \sqrt {-\Del }}{4\del }.
\end{eqnarray}
In this case, notice  that the
corresponding Riccati equation $v_\xi =a (1+v^2),\ a\in R^1$
 has the solutions
\[v(\xi )=\textrm{tan}(a\xi+\xi_0),\ v(\xi )=
- \textrm{cot}(a\xi+\xi_0)\]
with an arbitrary constant $\xi _0$.
Accordingly we obtain two explicit exact solutions for
the Kolmogorov-Petrovskii-Piskunov (\ref {KPP}) with $\del >0,\ \Del \le 0$:
\begin {eqnarray}
&&u'_1=-\frac \nu {2\del }+\vare _1\frac{\sqrt{-\Del }}{2\del }\textrm{tan}
(\vare _2 \frac{\sqrt{-\Del }}{2\sqrt{2\del }}x +\vare _1
\frac {\nu \sqrt{-\Del }}{4\del }t+\xi _0),
 \\ &&u'_2=-\frac \nu {2\del }-\vare _1\frac{\sqrt{-\Del }}{2\del }\textrm
{cot}(\vare _2 \frac{\sqrt{-\Del }}{2\sqrt{2\del }}
x +\vare _1
\frac {\nu \sqrt{-\Del }}{4\del }t+\xi _0),
\end {eqnarray}
where $\vare _1,\vare _2=\pm 1$ and $\xi _0$ is arbitrary. In the above
solutions, $\vare $ is again incorporated into $\vare _1$ due to the
same solution.

\section{Nonlinear interactions of traveling waves}

We make a Cole-Hopf transformation
\begin{equation}u=\al (\textrm{ln}f)_x=\al f_x/f,\ \al =\textrm{const.}
\end{equation}
for the Kolmogorov-Petrovskii-Piskunov equation (\ref{KPP}). We have
\begin{equation}(f_{xt}f-f_xf_t)f^2-(f_{xxx}f-3f_xf_{xx})f^2-2f_x^3f+
\mu f_xf^3+\nu \al f_x^2f^2+\del \al ^2 f_x^3f=0.
\end{equation}
Therefore if we choose $\al =\pm \sqrt{2/ \del  }=\vare \sqrt{2/ \del  },$
we get a bilinear equation
\begin{equation}f_{xt}f-f_xf_t-f_{xxx}f+3f_xf_{xx}+
\mu f_xf+\nu \al f_x^2=0.\label{bilineareq}
\end{equation}
After assuming a kind of solutions to be expressed by an exponential function
\[ f=A_1\textrm{e}^{k_1x+\omega _1t}
+A_2\textrm{e}^{k_2x+\omega _2t}+ A_3\textrm{e}^{k_3x+\omega _3t},\]
we find the conditions
 \begin{equation}\left\{
\begin{array}{l} 2k_i^3+\nu \al k_i^2+\mu k_i=0,\ 1\le i\le 3,\\
(\omega _i-\omega _j)(k_i-k_j)-(k_i^3+k_j^3-3k_ik_j^2-3k_i^2k_j)
\\ \quad +\mu (k_i+k_j) +2\nu \al k_ik_j=0,\
1\le i< j\le 3.
\end{array}\right.
\end{equation}
By solving this equation, we get a nontrivial solution to the equation
(\ref{bilineareq})
\begin{equation} f=A_1
+A_2\textrm{e}^{k_+x+(k_+^2-\mu)t}+ A_3\textrm{e}^{k_-x+(k_-^2-\mu)t},
\end{equation}
where $A_i,\ 1\le i\le 3$, are arbitrary constants and
\[ k_\pm =\frac{-\vare \nu \pm \sqrt{\Delta }}{2\sqrt{2\del }}
 =\frac{-\vare \nu \pm \sqrt{\nu ^2 -4\mu \del  }}{2\sqrt{2\del }}.
\]
Finally we obtain two different solutions to the
 Kolmogorov-Petrovskii-Piskunov equation
(\ref{KPP}) with $\del >0,\, \Del \ge 0$:
\begin{equation}
u_4(x,t)=u_4(x,t;A_1,A_2,A_3)=\sqrt{\frac{2}{\del }}\,\frac{A_2k_+\textrm{e}^{
\eta _+}+A_3k_-\textrm{e}^{\eta _-}}{A_1+A_2\textrm{e}^{\eta _+}+A_3\textrm{e}
^{\eta _-}},\label{solu4}
\end{equation}
with
\[ k_\pm =\frac{ -\nu \pm \sqrt{\Delta }} {2\sqrt{2\del }},\ \eta _\pm=
k_\pm x+(k_\pm ^2-\mu )t; \]
\begin{equation}
u_5(x,t)=u_5(x,t;A_1,A_2,A_3)=-\sqrt{\frac{2}{\del }}\,\frac{A_2k_+\textrm{e}^{
\eta _+}+A_3k_-\textrm{e}^{\eta _-}}{A_1+A_2\textrm{e}^{\eta _+}+A_3\textrm{e}
^{\eta _-}},\label{solu5}
\end{equation}
with
\[ k_\pm =\frac{\nu \pm \sqrt{\Delta } } {2\sqrt{2\del }}
,\ \eta _\pm=
k_\pm x+(k_\pm ^2-\mu )t;\]
where $A_1,A_2,A_3$ are three arbitrary constants.
We note that $u_4(-x,t,A_1,A_2, $ $A_3)=u_5(x,t,A_1,A_3,A_2)$.
 Therefore for the case of FitzHugh-Nagumo equation,
$u_5$ is exactly a solution lost in Ref. \cite{Kawahara}.
The solutions $u_4,u_5$ describe the coalescence of two traveling fronts
or two  singular traveling  waves of
the same sense. Direct numerical calculations of nonlinear interactions
 for the FitzHugh-Nagumo  case of (\ref{solu4}) were done in Ref.
\cite{Kawahara}.
These two solutions are analytic on the whole plane of $(x,t)$ when
$A_iA_j>0,\,i\ne j$, and they blow up at some points of $(x,t)$ when $
A_i,\,1\le i\le 3$, don't possess the same signs.

\section{
B\"acklund
 transformations}
We know there are three solutions
\[ \beta =0, \ \be = C_\pm =\frac {-\nu \pm \sqrt{\Delta }}{2\del }\]
to the equation
$\mu \be  +\nu \be ^2+\del \be ^3=0$, when $\Del =\nu ^2-4\mu \del \ge0$.
Make a linear transformation
 \begin{equation}
u=\al \tilde u+\be ,\ \al \ne 0,\ \be =0\ \textrm{or}\ C_\pm .\label{BT}
\end{equation}
This moment
\[ \begin{array}{ll} \mu u+\nu u^2+\del u^3 &=\al (\mu +2\be \nu +3\be ^2\del )
\tilde u+\al ^2 (\nu +3\be \del )\tilde u^2+\al ^3\del \tilde u^3\vspace{1mm}
\\ &=
\left \{\begin {array}{cl}
\al \mu \tilde u+\al ^2\nu \tilde u^2+\al ^3 \del \tilde u^3,
& \textrm{for}\ \be =0,\\
-\al (2\mu +\be \nu)\tilde u+\al ^2(\nu +3\be
\del )\tilde u^2+\al ^3\del \tilde u^3,
&\textrm{for}\ \be =C_\pm .
\end {array}\right.
\end{array}\]
and thus the Kolmogorov-Petrovskii-Piskunov equation (\ref{KPP})
is equivalent to the following  new Kolmogorov-Petrovskii-Piskunov equation
\begin{eqnarray}
 \tilde u_t-\tilde
 u_{xx}&=&-(\mu +2\be \nu +3\be ^2 \del )\tilde u-
\al (\nu +3\be \del )
\tilde u^2-\al ^2\del \tilde u^3\nonumber\\
 & :=&
-\tilde \mu\tilde u-\tilde \nu\tilde u^2-\tilde\del\tilde
 u^3.\label{NKPP}
\end{eqnarray}
A  direct calculation yields that
 \begin{equation}
\tilde \Delta :=\tilde \nu^2-\tilde \mu \tilde \del =
\left\{\begin{array}{cl}
\al ^2 \Del \ &\textrm{when}\ \be =0;\vspace{1mm}\\
\frac14\al ^2(\nu +\sqrt{\Delta})^2,\  &\textrm{when}\ \be =C_+;\vspace{1mm}\\
\frac14\al ^2(\nu -\sqrt{\Delta})^2,\  &\textrm{when}\ \be =C_-.
  \end{array}\right.
\end{equation}
We remark that a similar equation $\tilde u_t-\tilde u_{xx}+\mu
\tilde u-\nu \tilde u^2+\del \tilde u^3=0$
is generated under the mirror transformation $u=-\tilde u$, which possesses
the same property as the old equation (\ref {KPP}).

The transformation (\ref{BT}) maps the case: $\del >0,\,\Delta \ge0$
into the same case: $\tilde\del >0,\,\tilde\Delta \ge0$.
Therefore a new Kolmogorov-Petrovskii-Piskunov equation (\ref {NKPP})
 has also five explicit exact
 solutions defined by
 \[\tilde u_i(x,t)=u_i(x,t)\left.\right|_{\mu=\tilde \mu,\nu=\tilde \nu,
\del =\tilde
\del },\ 1\le i\le 5,\]
and further five new exact solutions to the old
 Kolmogorov-Petrovskii-Piskunov equation (\ref{KPP}) may be presented
by $\al \tilde u_i(x,t)+\be ,\,1\le i\le5$.
However this transformation process hasn't given new kind of solutions to
(\ref{KPP}) beginning with the obtained five solutions, which will be
shown below.

Note that we have for $\al >0,\ \nu\ge 0$:
\[ \begin{array}{ll}
\sqrt{\tilde \Del }=\frac12\al  (\nu +\sqrt{\Del }),\ \tilde \nu -\sqrt{
\tilde \Del }=\al (- \nu + \sqrt{\Del }),\ \tilde \nu +\sqrt{
\tilde \Del }=2\al \sqrt{\Del },\  &\textrm{when}\ \be =C_+;\\
\sqrt{\tilde \Del }=\frac12\al  (\nu -\sqrt{\Del }),\ \tilde \nu -\sqrt{
\tilde \Del }=-\al (\nu +\sqrt{\Del }),\ \tilde \nu +\sqrt{
\tilde \Del }=-2\al \sqrt{\Del },\  &\textrm{when}\ \be =C_-,\ \mu \ge 0;\\
\sqrt{\tilde \Del }=\frac12\al  (\sqrt{\Del }-\nu ),\ \tilde \nu -\sqrt{
\tilde \Del }=-2\al \sqrt{\Del },\ \tilde \nu +\sqrt{
\tilde \Del }=-\al (\nu +\sqrt{\Del }),\  &\textrm{when}\ \be =C_-,\ \mu \le 0.
\end{array} \]
The concrete results of the B\"acklund
transformation
\begin{equation}
(\textrm{BT})_{\be }: u(x,t)\mapsto \alpha u(x,t)
\left.\right|_{\mu=\tilde \mu,\nu=\tilde \nu,
\del =\tilde
\del }+\be
\end{equation}
may be given out for $\al >0,\ \nu \ge0$ as follows:
\[ \begin{array}{ll}
(\textrm{BT})_{C_+}:\ &
u_1(x,t;\vare _1,\vare _4)\mapsto u_3(x,t;\vare _3
=-\vare _1,\vare _6=-\vare _4),\\
&
u_2(x,t;\vare _2,\vare _5)\mapsto u_2(x,t;-\vare _2,-\vare _5),\\
&
u_3(x,t;\vare _3,\vare _6)\mapsto u_1(x,t;\vare _1=-\vare _3,\vare _4
=\vare _6),\\
&
u_4(x,t;A _1,A _2,A_3)\mapsto u_4(x,t;A_2,A_1,A_3),\\
&
u_5(x,t;A _1,A _2,A_3)\mapsto u_5(x,t;A_3,A_2,A_1);
\end{array}\]
\[ \begin{array}{ll}
(\textrm{BT})_{C_-,\mu \ge0}:\ &
u_1(x,t;\vare _1,\vare _4)\mapsto u_2(x,t;\vare _2
=-\vare _1,\vare _5=\vare _4),\\
&
u_2(x,t;\vare _2,\vare _5)\mapsto u_3(x,t;\vare _3
=-\vare _2,\vare _6=-\vare _5),\\
&
u_3(x,t;\vare _3,\vare _6)\mapsto u_1(x,t;\vare _1
=\vare _3,\vare _4=-\vare _6),\\
&
u_4(x,t;A _1,A _2,A_3)\mapsto u_4(x,t;A_2,A_1,A_3),\\
&
u_5(x,t;A _1,A _2,A_3)\mapsto u_5(x,t;A_3,A_2,A_1);
\end{array} \]
\[ \begin{array}{ll}
(\textrm{BT})_{C_-,\mu \le0}:\ &
u_1(x,t;\vare _1,\vare _4)\mapsto u_2(x,t;\vare _2
=\vare _1,\vare _5=-\vare _4),\\
&
u_2(x,t;\vare _2,\vare _5)\mapsto u_1(x,t;\vare _1
=\vare _2,\vare _4=-\vare _5),\\
&
u_3(x,t;\vare _3,\vare _6)\mapsto u_3(x,t;-\vare _3,-\vare _6),\\
&
u_4(x,t;A _1,A _2,A_3)\mapsto u_4(x,t;A_3,A_2,A_1),\\
&
u_5(x,t;A _1,A _2,A_3)\mapsto u_5(x,t;A_2,A_1,A_3).
\end{array} \]
When $\be =0$, the transformation (\ref{BT}) also
maps the other case: $\del >0,\ \Del \le 0$ into the same case: $
\tilde \del >0,\ \tilde \Del \le 0$.
We may show that
 $(\textrm{BT})_{\be =0,\al >0}$ is an identity map on the
set of solutions $\textrm{Span}\{u_i,u'_j|\,1\le i\le 5,j=1,2\}$ and that
\[\begin{array}{ll}
(\textrm{BT})_{\be =0,\al <0}:
&u_1(x,t;\vare _1,\vare _4)
 \mapsto u_1(x,t;-\vare _1,\vare _4),\\
&u_2(x,t;\vare _2,\vare _5)
 \mapsto u_3(x,t;\vare _3=\vare _2,\vare _6=-\vare _5),\\
&u_3(x,t;\vare _3,\vare _6)
 \mapsto u_2(x,t;\vare _2=\vare _3,\vare _5=-\vare _6),\\
&u_4(x,t;A_1,A_2,A_3)
 \mapsto u_5(x,t;A_1,A_2,A_3),\\
&u_5(x,t;A_1,A_2,A_3)
 \mapsto u_4(x,t;A_1,A_2,A_3),\\
&u'_1(x,t;\vare _1,\vare _2)
 \mapsto u'_1(x,t;-\vare _1,\vare _2),\\ & u'_2(x,t;\vare _1,\vare _2)
 \mapsto u'_2(x,t;-\vare _1,\vare _2).
\end{array}\]
The rest case of $(\textrm{BT})_{C_\pm}$ may be computed similarly.
It is interesting to note
 that $(\textrm{BT})_{\be =0,\al <0}$ casts the solution (\ref{solu4})
into the solution (\ref{solu5}) and vice verse.

We point out that we may also transform a more general equation
\begin{equation}
w_t-w_{xx}=f(w)=a +bw+cw^2+dw^3,\ a\ne 0,\label{nonzeroKPP}
\end{equation}
where $a,b,c,d$ are real constants, into the
 Kolmogorov-Petrovskii-Piskunov equation (\ref{KPP})
under the linear transformation $w=\alpha u+\beta ,\ \alpha \, , \beta
=\textrm{consts.}$ Therefore we can generate solutions to (\ref{nonzeroKPP})
by the obtained results. However we should note that the equation
(\ref{nonzeroKPP}) lost the property $f(0)=0$.

\section{Discussions}

The Kolmogorov-Petrovskii-Piskunov equation (\ref{KPP}) contains the following
various equations with $\mu +\nu +\del =0$, for which  there is
always the condition $\Del =\nu ^2-4\mu \del
=(\mu -\del )^2\ge 0$ and thus has five explicit solutions.

(i) The nonintegrable Newell-Whitehead \cite{Newell}
equation
\begin{equation}
u_t-u_{xx}=u-u^3
\end{equation}
is a special case with $\mu =-1,\ \nu=0,\ \del =1$.
 The cases of $\mu =0,\ \nu=-1,\ \del =1$ and
$\mu =1,\ \nu=-2,\ \del =1$
 engender the
equations
\begin{eqnarray}& &
u_t-u_{xx}=u^2(1-u),\\ & &
u_t-u_{xx}=-u(1-u)^2,
\end{eqnarray}
respectively. The above three equations are all simple
 generalizations of the Fisher
equation.
Interestingly, an another simple generalization of the Fisher equation
$u_t-u_{xx}=u(1-u)^2$ has no nonconstant solution of the form
\[u=\sum_{i=0}^Ma_i\textrm{tanh}^i(kx-\omega t+\xi _0),\
u=\sum_{i=0}^Mb_i\textrm{coth}^i(kx-\omega t+\xi _0), \ a_i,\,b_i\in R^1.\]
This is in agreement with the result in Ref. \cite{Cohen}, where
it was shown that this equation has no solution of the form
$a_1\textrm{tanh}(kx-\omega t+\xi _0)+a_0$.

(ii) The case of $\mu =a,\ \nu=-(a+1),\ \del =1$ engenders
the FitzHugh-Nagumo equation
\begin{equation}
u_t-u_{xx}=u(u-a)(1-u),
\end{equation}
where $a$ is arbitrary.
This equation may describe nerve pulse propagation in nerve fibers
and wall motion in liquid crystals.
 Its solutions were discussed in \cite{Hereman}
\cite{Kawahara}.

Our method in Section 2
is a kind of combination of the direct method \cite{WangKL}
 and
the ansatz method \cite {Lu} \cite{Yang} and thus we term
it the combined ansatz method.
The idea is to make the unknown variable $u$ to be a practicable
function $g(v)$ of the ansatz unknown variable $v$ which
satisfies a differential equation solvable by quadratures.
This allows us to solve
a large class of physically important nonlinear equations
including some nonintegrable ones, for example,
2$D$-KdV-Burgers equation \cite{Parkes} and 7th generalized KdV equation
\cite{Ma}.
 The crucial point is to choose the proper ansatz equations
 solvable by quadratures. We here list two useful ansatz equations and
their solutions.
The first one is the Bernoulli equation
\begin{equation}v_\xi =a v+b  v^\al ,\ a,b,\al \in R^1,\ ab\ne 0, \ \al \ne 1.
\label{Bernoulli}
\end{equation}
It has a general solution
\begin{equation}
v=\left [-\frac a b \frac 1
 {\xi _0\textrm{e}^{a(1-\al )\xi}+1}\right]^{\frac 1{\al -1}}=\left
 \{\begin{array}{cl}
(-\frac a{2b})^{\frac 1{\al -1}}, &\textrm{for}\ \xi_0=0,\vspace{1mm}\\
\{-\frac a{2b}[ \textrm{tanh}(\frac {a(\al -1) }2\xi -\frac {\textrm{ln}\xi_0}
2)+1]\}^{\frac 1{\al -1}}, &\textrm{for}\ \xi _0>0,\vspace{1mm} \\
\{-\frac a{2b} [ \textrm{coth}(\frac {a(\al -1) }2\xi -\frac {\textrm{ln}
(-\xi _0)}
2)+1]\}^{\frac 1{\al -1}}, & \textrm{for}\ \xi _0<0,
\end{array}
\right.
\end{equation}
where $\xi _0$ is arbitrary.
The second one is the Riccati equation
\begin{equation}
v_\xi =a_0+a_1v+a_2v^2,\ a_i\in R^1,\ a_2\ne 0.\label{Riccati}
\end{equation}
This equation has the following solutions
\begin{eqnarray}
v&=&-\frac {a_1}{2a_2},\  -\frac 1 {a_2\xi +\xi _0}-\frac {a_1}{2a_2},
\ (\Del =0);\\
v&=&-\frac {\vare \sqrt{\Del }}{a_2}\frac 1 {\xi _0\textrm{exp}({-\vare
\sqrt{\Del }\xi })+1}
+\frac {\vare \sqrt{ \Del }}{2a_2}-\frac {a_1}{2a_2}\ \  (\vare =\pm 1),
\nonumber\\
 &= & \left \{\begin {array}{cl}
\frac {\vare \sqrt{ \Del} }{2a_2}-\frac {a_1}{2a_2},
&\textrm{for}\  \xi _0=0,
\vspace{1mm}
\\-\frac { \sqrt{\Del }}{2a_2}\textrm{tanh}(\frac{
\sqrt{\Del} }{2}\xi -\frac {\vare \textrm{ln}\xi _0}2)-\frac {a_1}{2a_2},
&\textrm{for}
\ \xi _0>0,\vspace{1mm}
\\
 -\frac { \sqrt{\Del }}{2a_2}\textrm{coth}(\frac{
\sqrt{\Del} }{2}\xi -\frac {\vare \textrm{ln}(-\xi _0)}2)-\frac {a_1}{2a_2},
 &\textrm{for}\ \xi _0<0,
\end{array}\right.\ (\Del >0);\quad \\v& =&
\left \{\begin {array}{l}
\frac { \sqrt{-\Del} }{2a_2}\textrm{tan}(\frac{
\sqrt{-\Del} }{2}\xi +\xi _0)-\frac {a_1}{2a_2},\vspace{1mm}\\
-\frac { \sqrt{-\Del} }{2a_2}\textrm{cot}(\frac{
\sqrt{-\Del} }{2}\xi +\xi _0)-\frac {a_1}{2a_2},
\end{array}\right. \ \ (\Del <0);
\end{eqnarray}
where $\Del =a_1^2-4a_0a_2$ and  $\xi_0$ is arbitrary.
The Bernoulli equation and the Riccati equation have been
appeared in \cite{Lu}
\cite {Yang}. However, singular solutions
 of these two equations are all missing in their works.

For the general Fisher equation
\begin{equation}
u_t-u_{xx}+\mu u+\nu u^2=0,\  \mu,\,\nu \in R^1,\label{Fisher}
\end{equation}
the same combined ans\"atze as ones
 for the Kolmogorov-Petrovskii-Piskunov (\ref{KPP})
may also result in solutions, but only a kind of solutions
\begin{equation}
u(x,t)=-\frac \mu {2\nu} -\frac {|\mu |}{4\nu }+\frac {\vare _1\mu }{2\nu }v(
\frac {\vare _2 \sqrt{6|\mu|}}{12}x+\frac {5\vare _1\mu }{12}t)+
\frac {|\mu |}{4\nu }v^2(
\frac {\vare _2 \sqrt{6|\mu|}}{12}x+\frac {5\vare _1\mu }{12}t),\label{fsolu}
\end{equation}
where $v_\xi =1-v^2,\ \vare _1,\vare _2 =\pm1 $.
Particularly,  the Fisher elution (\ref{Fisher}) hasn't any
solution of the form
\begin{equation}
u=\sum_{i=0}^Ma_i\textrm{tan}^i(kx-\omega t+\xi _0),\
u=\sum_{i=0}^Mb_i\textrm{cot}^i(kx-\omega t+\xi _0),\ a_i,\,b_i\in R^1.
\end{equation}
It is worth pointing out that the Fisher equation can't  be solved  through
the ansatz method proposed in Refs. \cite{Lu} \cite {Yang}.
When we choose $v=\textrm{tanh}(\xi +\xi _0)$ or $v=\textrm{coth}
(\xi +\xi _0)$,
the solution (\ref{fsolu}) yields two explicit solutions, for instance,
a traveling front solution
\begin{equation}
u(x,t)=-\frac \mu {2\nu} -\frac {|\mu |}{4\nu }+\frac {\vare _1\mu }{2\nu }
\textrm{tanh}(
\frac { \sqrt{6|\mu|}}{12}x+\frac {5\vare _1\mu }{12}t+\xi _0)+
\frac {|\mu |}{4\nu }
\textrm{tanh}^2(
\frac { \sqrt{6|\mu|}}{12}x+\frac {5\vare _1\mu }{12}t+\xi _0).\label{fsolu1}
\end{equation}
They contain two solutions
 of the standard Fisher
equation $u_t-u_{xx}=u(1-u)$.

Finally we remark that any Riccati equation possesses  an important property:
given a particular solution, its general solution may be found by quadratures.
This property is named the Riccati property by Fuchssteiner and Carillo \cite
{FuchssteinerC} and a method to construct ordinary differential equations
which enjoy the Riccati property is proposed in their works
 \cite{FuchssteinerC} \cite {CarilloF}. We may also take the differential
equations proposed in \cite {FuchssteinerC} \cite {CarilloF} as the basic
 ansatz equations. This may make more nonlinear equations to be solvable by
quadratures.

\vskip 5mm

\noindent{\bf Acknowledgment:} One of the authors (W. X. Ma) acknowledges
financial support from Alexander von Humboldt Foundation.

\end{document}